\documentclass{bmvc2k}
\usepackage{graphicx}
\usepackage{bbm}
\usepackage{amsmath}
\usepackage{amsfonts}
\usepackage{multirow}
\usepackage{hyperref}
\usepackage{wrapfig}

\title{Mutual Contrastive Low-rank Learning to Disentangle Whole Slide Image Representations for Glioma Grading}

\addauthor{Lipei Zhang}{lz452@cam.ac.uk}{1}
\addauthor{Yiran Wei}{yw500@cam.ac.uk}{2}
\addauthor{Ying Fu}{fuying@bit.edu.cn}{3}
\addauthor{Stephen Price}{sjp58@cam.ac.uk}{2}
\addauthor{Carola-Bibiane Schönlieb}{cbs31@cam.ac.uk}{1}

\addauthor{Chao Li (corresponding author)}{cl647@cam.ac.uk}{1,2}

\addinstitution{
 Department of Applied Mathematics \\
 and Theoretical Physics \\
 University of Cambridge
}
\addinstitution{
 Department of Clinical Neuroscience\\
 University of Cambridge
}
\addinstitution{
 School of Computer Science and Technology\\
 Beijing Institute of Technology
}

\runninghead{Zhang. L. et al.}{Mutual Contrastive Low-rank Learning}

\begin{document}

\maketitle

\begin{abstract}
Whole slide images (WSI) provide valuable phenotypic information for histological assessment and malignancy grading of tumors. The WSI-based grading promises to provide rapid diagnostic support and facilitate digital health. Currently, the most commonly used WSIs are derived from formalin-fixed paraffin-embedded (FFPE) and Frozen section. The majority of automatic tumor grading models are developed based on FFPE sections, which could be affected by the artifacts introduced by tissue processing. The frozen section exists problems such as low quality that might influence training within single modality as well. To overcome this problem in a single modal training and achieve better multi-modal and discriminative representation disentanglement in brain tumor, we propose a mutual contrastive low-rank learning (MCL) scheme to integrate FFPE and frozen sections for glioma grading. We first design a mutual learning scheme to jointly optimize the model training based on FFPE and frozen sections. In this proposed scheme, we design a normalized modality contrastive loss (NMC-loss), which could promote to disentangle multi-modality complementary representation of FFPE and frozen sections from the same patient. To reduce intra-class variance, and increase inter-class margin at intra- and inter-patient levels, we conduct a low-rank (LR) loss. Our experiments show that the proposed scheme achieves better performance than the model trained based on each single modality or mixed modalities and even improves the feature extraction in classical attention-based multiple instances learning methods (MIL). The combination of NMC-loss and low-rank loss outperforms other typical contrastive loss functions. The source code is in \url{https://github.com/uceclz0/MCL_glioma_grading}.
\end{abstract}

\section{Introduction}
Glioma is the most frequent malignant primary brain tumor, characterized by remarkable infiltration and tumor heterogeneity \cite{grade_glio,who2007,who2016}. According to the World Health Organization (WHO) classification, glioma is classified into four grades, where grade IV represents the most aggressive type, and lower grade glioma (LGG), i.e., grades II and III, are less aggressive \cite{who2007,who2016}. Glioma grading has crucial significance for treatment planning and risk stratification towards precision medicine \cite{chao_nature,chao_r_o,chao_neo}. The current practice of glioma grading is based on the histology assessment of tumor specimens, which is time-consuming and requires high professional expertise. Therefore, an accurate and automatic approach to glioma grading based on the WSIs promises to provide rapid diagnostic support for timely clinical decision-making. Furthermore, a computer-assisted method could help facilitate digital health and enhance the accessibility of medical resources.

The FFPE tissue section is generally used as the diagnostic standard in clinical practice, which can be used for long-term storage because of formalin-fixed paraffin-embedded. Due to the gigapixel size of the images and the complexity of tumor tissue, splitting WSI into small patches is used as an effective method in model training \cite{sensors,neuro_oncology_advance}. By using this operation, previous studies proposed machine learning approaches based on feature engineering \cite{wang,mil_provement}. Although providing reasonable performance, these approaches were prone to model generalizability due to the less robust features extracted from diverse tumor tissue. Recently, most state-of-the-art models employed the deep transfer learning approach to transfer the pre-trained weights from ImageNet \cite{sensors,imagenet_cvpr09,pei,neuro_oncology_advance}. These deep learning methods usually neglect the correlation among different instances because patches are typically described by weak annotation such as a global label. Therefore, some researchers proposed some attention-based multiple instance learning (MIL) methods by integrating all patches from one WSI \cite{amil,transmil,clam}. However, all previous studies only consider extracting features or classification on FFPE sections. These studies might be affected by the bias of FFPE tissue. Moreover, a single section modality may not facilitate learning relevant image representations for tumor grading. Importantly, the artifacts introduced by formalin could affect the interpretation of histological specimens \cite{ffpe_artifact}. Specifically, the procedure of prefixation and fixation could influence the morphological quality of FFPE specimens \cite{fixation}. These artifacts could pose particular challenges to the model training based on WSI. In parallel, another modality of WSI from frozen tissue procedure provides a rapid approach to tumor grading to guide intra- or peri-operative clinical decisions. Although frozen sections typically contain limited tissue, the sample hydration and cellular morphology of the frozen tissue can be preserved at a natural state \cite{frozen_adv}, which may provide crucial for tumor grading. However, the frozen sections also exist that some potential problems, such as poor quality, influence the performance \cite{frozen_cls}. Hence, adopting two modalities could promote extracting significantly complementary information for tumor grading by a multi-modality training scheme.

The rising mutual learning scheme promises to jointly optimise WSI representations across FFPE and frozen sections. Mutual learning \cite{deep_mutual_learning} was initially proposed to facilitate direct knowledge distillation based on joint training scheme. Later, a hierarchical architecture with multiple classifier heads was proposed to improve model generalisation \cite{collaborative_2018}. Additionally, peer mutual learning was proposed for online unified knowledge distillation \cite{peer_mutual}. Mutual learning was also successful in classification tasks based on audio-video data \cite{audio_video}. However, the classic losses used in mutual learning may not be able to extract extra complementary information from the latent spaces of multi-modality. On the other hand, contrastive loss \cite{simclr,mocov2} has shown the capacity of extracting complementary representations from latent vectors. Therefore, introducing a contrastive loss into mutual learning schemes could allow learning complementary representations jointly from multiple modalities with better generalisation, such as joint learning on visual-textual information \cite{multimodal_cvpr_2021} and predicting isocitrate dehydrogenase mutation of glioma \cite{yi}. Here we hypothesise that a mutual contrastive learning scheme could achieve better performance in tumour grading based on FFPE and frozen sections. 


However, the contrastive loss is only about the variance of inter-modality rather than intra-class. Based on the diversity characteristic of brain tumors, cancer cells are genetically aberrant and can be divided into different sub-types at each grade \cite{diversity}, and there are relatively similarities between different grades \cite{similarity_cell}, which causes intra-class variance and decreases the inter-class margin. There are potential solutions that aim at solving these problems, such as pairwise or triplet losses \cite{triplet_loss}. As a result, they carry an extra computation and learning burden in selecting and computing multiple pairs or triplets. In addition, the noised images will introduce uncertainty in the training phase as well. Therefore, some researchers adapted low-rank constraint to explicitly improve the discriminative capacity on natural images without specific pairing and reduce uncertainty from noised data. The low-rank constraint is achieved via a linear transformation enforcing the minimum rank of each class feature sub-matrix, and an orthogonalization constraint on the matrix of features of all classes \cite{low_rank_early,ole}. Meanwhile, unlike the well-known cross-entropy loss computed on each paired data vectors individually, low rank is able to globally optimize the lowest-rank representation on a collection of vectors, which will be more robust for noised data \cite{low_rank_seg}. Based on these methods, a multi-modality mutual contrastive low-rank learning scheme becomes promising, which could simultaneously achieve disentangling representation from multi-modal and reducing intra-class variance, increase inter-class margin at intra- and inter-patient levels. 

In our paper, we first hypothesize that integrating FFPE and frozen sections could train a more robust model to learn the high-level representations reflecting tumor malignancy, with less bias from the artefacts caused by tissue processing and low quality. To achieve this goal, we design a parallel mutual learning scheme to facilitate the integration of FFPE and frozen sections in model training. In this scheme, we design a normalised multi-modal contrastive loss (NMC loss) to disentangle representation with the sphere projection \cite{sphere}. Meanwhile, we adopt low-rank loss to promote latent vectors from the same class to lie in a linear sub-space by lowering the matrix rank and latent vectors at inter-class being in an orthogonal sub-space, which achieves
better discriminative representation disentanglement at the intra- and inter-patient levels.

To our best knowledge, this is the first multi-modality mutual contrastive learning approach for glioma grading in the field of digital pathology. Our contributions include:
 
 - a mutual contrastive low-rank learning (MCL) scheme for joint optimization of model training based on the WSIs of FFPE and frozen sections.
 
 - an NMC loss to improve the ability to disentangle multi-modality representations in the mutual learning process.
 
 - a low-rank loss to reduce intra-class variance, and increase inter-class margin at intra- and inter-patient levels.

\section{Methods} 
\begin{figure*}[ht]
    \centering
    \includegraphics[scale=0.25]{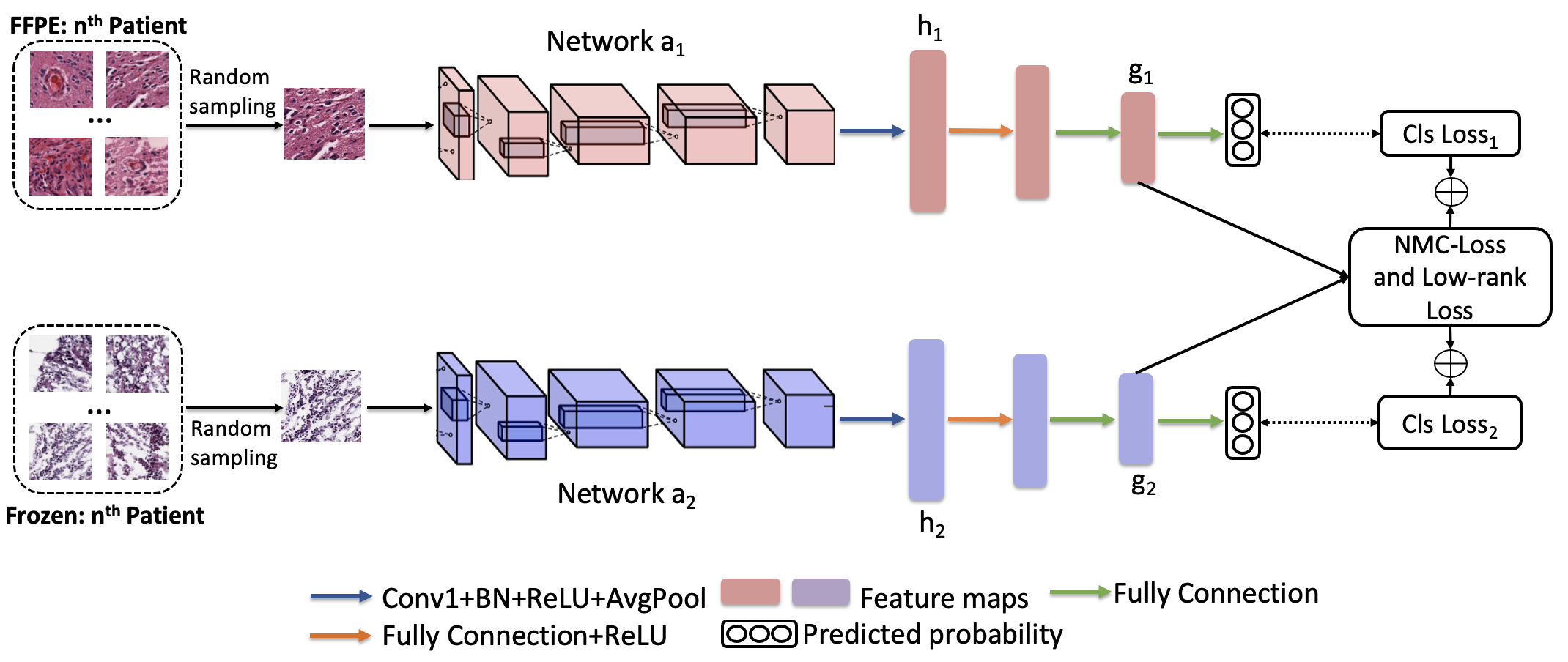}
    \caption{The pipeline of gliomas grading on FFPE sections and frozen sections. The FFPE and frozen patches will be randomly sampled from the bags from the same patient. Two sorts of patches are input into two networks without weight-sharing. The feature vectors from different layers play different roles in the representation disentanglement and classification.} 
\end{figure*}

The proposed parallel mutual optimization learning scheme is shown in Fig. 1. Initially, a random sampling strategy is adopted for the mutual training. The FFPE and frozen patches will be randomly sampled from the bags of the same patient. The paired images are input into the Network $\alpha_1$ and $\alpha_2$, respectively. In each branch, the set of feature vectors ($h_1$ and $h_2$) after the first fully connected layer and ReLU are input into the next two fully connected layers respectively for multi-modal representation disentanglement and low-rank optimization. Combining contrastive loss, low-rank loss and classification loss could optimize each modality network jointly via the back-propagation. We introduce details of the main design in the following parts. 


\textbf{Non-linear representation}. We adopt three fully connections with ReLU in our scheme to promote better non-linear projection of latent vectors for representation learning. In each modal network, the latent vectors after average pooling are transformed by first fully connection, ReLU and second hidden layer with the same dimensional transformation, so that we can obtain a representative vector with the non-linear projection $(z_i = g(h_i) = W^{2}\sigma(W^{1}h_i))$, where $\sigma$ is s a ReLU non-linearity. Referred by the experiments of SimCLR \cite{simclr}, a non-linear operator can remove variant information, e.g., the color or orientation of objects resulting from various staining procedures from multiple centers. Therefore, these latent vectors can be more suitable for representation disentanglement and the final classification head can leverage the nonlinear transformation to maintain more useful information in $h$, which could boost the performance. 

\textbf{NMC-loss}. For this design, formally given a mini-batch of size N, we firstly view the FFPE {$x^{a}_{1},...,x^{a}_{K}$} and the frozen section , {$x^{b}_{1},...,x^{b}_{K}$} images as images sampled from different augmented views on the same patient. There are $2N-1$ pairs totally, among which we can regard the corresponding augmented sample $x^{b}_{i}$ as a positive pair {$x^{a}_{k},x^{b}_{k}$} and other $2N-2$ pairs are negative samples. In standard contrastive loss definition presented in \cite{simclr,mocov2}, $l2$ normalization was used for reference and augmented images. However, the image pairs used in self-supervised learning keep unified features and distribution, and $l2$ normalization can scale latent vector into a valid range. In our task, the image pairs include two domains with different distributions. Therefore, we adopt layer normalization, as shown in Eq. 1, to center and rescale the latent space into the same sphere, which can improve the efficiency of disentanglement.

\begin{equation}
    \hat{g}_{n} = \frac{g_n - \mu_n}{\sqrt{(\sigma_n)^2 + \epsilon}}
\end{equation}
where $n \in \{1, 2\}$ and $g_n$ denotes the latent vector from FFPE or frozen section. $\mu$ and $\sigma$ are the mean and variance of each batch. Moreover, we denote $sim(\hat{g}_1, \hat{g_2}) = \frac{\hat{g}_1^T \hat{g_2}}{||\hat{g}_1||\cdot||\hat{g_2}||}$ as the cosine similarity between $\hat{g}_1$ and $\hat{g_2}$. The NMC-loss function can be defined as:

\begin{equation}
    L^{a}_{k} = -log \frac{exp(sim(\hat{g}^{a}_{k},\hat{g}^{b}_{k})/\tau)}{\sum_{i\in I}\mathbbm{1}_{i \neq k} exp(sim(\hat{g}^{a}_{k},\hat{g}^{b}_{i})/\tau)}
\end{equation}

where $\mathbbm{1}_{i \neq k} \in \{0, 1\}$ is an indicator, which values $\mathbbm{1}$ only when $i \neq k$. We also define $\tau$ as a temperature hyper-parameter. To identify all positive pairs in this batch, the NMC-loss is further defined as ($L^{b}_k$ follows the same calculation with $L^{a}_k$) :

\begin{equation}
    L_{nmc} = \frac{1}{2N}\sum_{k=1}^K (L^{a}_{k} + L^{b}_{k})
\end{equation}

\textbf{Low-rank loss}. Inspired by the idea of low rank of representations \cite{ole,low_rank_seg}, we can consider that successful training of the network would result in the classifier vectors remaining orthogonal at the end if non-linear representation $g_n$ can be in the orthant and low-rank. More precisely, we can consider a feature embedding of each modality $X^a = [x^{a}_{1}|x^{a}_{2}|...|x^{a}_{N}]$, where each column $x^{a}_{i} \in R^d, i = 1,....,N$, $a \in [1, 2]$ and $|$ represents vertical concatenation. The $X^a$ is obtained from a given training samples $Y$ with minibatch size $N$, $X = \phi(Y;\theta)$, and $X$ is the $N \times D$ deep embedding from extractor $\phi$ with parameter $\theta$. We further assume that $X^{a}_c, Y^{a}_c$ are the sub-feature matrices and input 5data respectively belonging to grade $c$ and modality $a$. To achieve better discriminative representation disentanglement in intra- and inter-patient levels, the sub-matrices from two modalities will be concatenated from a vertical direction such as $M = [X^{1}|X^{2}]$. In each minibatch, our low-rank loss can be defined as the following equations: 

\begin{equation}
    \begin{split}
    L_{lr} & = \sum_{c=1}^{C} max(\Delta, ||M_c||_{*}) - ||M||_*\\
    &  = \sum_{c=1}^{C} max(\Delta, ||[\phi(Y^{1}_c;\theta)|\phi(Y^{2}_c;\theta)]||_{*}) - ||[\phi(Y^{1};\theta)|\phi(Y^{2};\theta)]||_*
    \end{split}   
\end{equation}

Where $||.||_*$ means the matrix nuclear norm (the sum of the singular values) $\Delta \in R$ denotes a bound on the intra-class nuclear loss that can avoid the training collapse resulting from feature value to zero. In our experiments, we set $\Delta=1$. The first term in loss function can minimize the rank of each grade feature subspace and the second term promotes inter-class to be linearly orthogonal. 

To further clarify the optimization in backpropagation, we can calculate a simplified subgradient of the nuclear norm. Based on the SVD decomposition and deduction from \cite{ole}, the descent direction can be defined as the following equation:
\begin{equation}
    \begin{split}
    gL_{lr}(M) & = \sum_{c=1}^{C} [Z^{(l)}_c|U_{c1}V^{T}_{c1}|Z^{(r)}_c] - U_1V^{T}_1\\\
    \end{split}   
\end{equation}

Where, $Z^{(l)}_c$ and $Z^{(r)}_c$ denote fill matrices of zeros to keep the original dimension of $M$. $U_1$ and $V_1$ are the principal left and right singular vectors from $M$ and $U_{c1}$ and $V_{c1}$ are chosen by the singular value being greater than a fixed threshold.

\textbf{Model optimization}. Our learning scheme consists of two functions to achieve joint optimization in two classification tasks. Each function consists of a cross-entropy (CE) loss with Taylor Softmax \cite{taylor_ce}, and a NMC-loss and low rank loss. The Taylor Softmax CE can smooth labels to reduce over-fitting and it already has been proved effectiveness in many competition solutions. This loss can be formed as Eq. 3 and the total loss function of each modality can be expressed as Eq. 4:
\begin{equation}
    \begin{split}
        & L_{cls}(f(x), y) = \sum_{i=1}^{t} \frac{(1-f_y(x))^i}{i} \\
    \end{split}
\end{equation}

\begin{equation}
    \begin{split}
        & L_{FFPE} = L_{cls_1} + L_{nmc} + L_{lr}\\
        & L_{Frozen} = L_{cls_2} + L_{ins} + L_{lr}
    \end{split}
\end{equation}

where the $f_y(x)$ denotes the y-th element of $f(x)$ and $f(.)$ is a CNN with the classification layer. $t$ is the term number of the Taylor series and we set $t=3$ as the same as the setting from the original paper.

\section{Experimental Setup}
\subsection{Datasets}
We utilized the WSI of glioblastoma (GBM) and LGG from the Cancer Genome Atlas (TCGA) dataset \cite{tcga}, with clinical details and Hematoxylin and Eosin (HE) stained sections available. We only selected 499 patients (108 grade II, 94 grade III, and 297 grade IV) with both FFPE and frozen sections available. 

For data pre-processing, we designed three steps: 1) transforming a low-dimension version of WSI into HSV color space and separating HE-stained tissue from the background using Otsu’s Binarization on the saturation channel \cite{hsv_seg}; 2) patching a number of non-overlapping $500\times 500$ instance-level images at $20\times$ magnification; 3) a blob detection procedure \cite{blob_detection} to further remove redundant patches containing insufficient tissue. The numbers of finally included patches were 1,680,714 for FFPE sections and 483,886 for frozen sections. 

To evaluate the proposed scheme, the dataset was randomly divided into 319 patients for the training set, 80 patients for the validation set and 100 patients for the testing set. In testing set, it included 27 grade II, 25 grade III, and 48 grade IV. Moreover, to increase sample size, we cropped sub-regions of patches into a size of $224 \times 224$. In addition, we applied data augmentation techniques (random rotation of 90°, 180°, 270°, random flipping image along axis, shift hue saturation value and brightness contrast) to increase the training sample size.

\subsection{Training Details}
The training environment was based on PyTorch 1.6.0 backend with acceleration by Nvidia RTX 3090. The batch size was set to 32, corresponding to 32 pairs of FFPE and frozen images in the mutual training, while 32 of FFPE or frozen images in single training and mixed training. The input was $224 \times 224 \times 3$. We used the loss function described in. Eq.4 for our experiments. The number of training epochs was $10$, and the optimizer was Adam with default parameters. Cosine annealing warm restarts were adopted with an initial learning rate of $1.6 \times 10^{-4}$. 

We trained different CNN backbones with single-modal input training (baseline) \cite{sensors,neuro_oncology_advance}, mixed-modal training and our proposed mutual training scheme. Meanwhile, some comparisons with state of art methods (SOTAs), such as attention MIL (A-MIL) \cite{amil}, TransMIL \cite{transmil} and CLAM \cite{clam}, were introduced. In the original papers, the feature vector of each patch was from the backbone with ImageNet pre-trained weights. Therefore, we further used the trained weights from the different learning schemes to extract feature vectors at patch-level. Moreover, we compared the performance of different metrics loss in mutual training such as Kullback-Leibler divergence \cite{deep_mutual_learning}, marginal triplet loss \cite{triplet}, NT-logistic loss, NT-loss \cite{simclr} and angular margin contrastive loss \cite{amc_loss} as well.

\subsection{Testing Details}
\begin{figure*}[ht]
    \centering
    \includegraphics[scale=0.3]{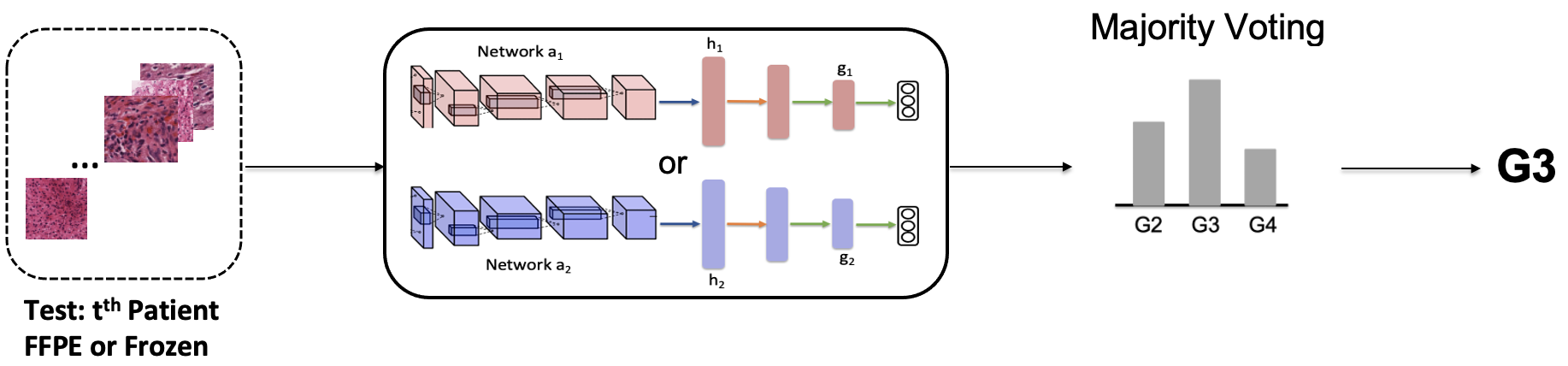}
    \caption{The pipeline of gliomas grading on FFPE section and frozen section.} 
\end{figure*}
The details of the inference phase are shown in Fig. 2. FFPE and frozen sections were classified separately by Network $a_1$ or $a_2$ with respective classification heads. After predicting all images in a patient's bag, the number of predicted grades was counted by the histogram and majority voting was used to determine the final predicted tumor grade. Moreover, in the comparison study, we also followed protocols in A-MIL \cite{amil}, CLAM \cite{clam} and TransMIL \cite{transmil} to train and test at patient-level feature matrices.

\begin{table}[htbp]
    \renewcommand{\arraystretch}{1.3}
    \centering
    \caption{Comparison with different learning schemes. Single training only trained and tested on the same single modality. Mixed training mean combining modalities for training on one model and separately testing in each. Mutual training followed proposed method to train and test. The ImageNet pre-trained weights, single trained weights, mixed trained weights and mutual trained weights were used in feature extraction and combined with A-MIL, TransMIL and CLAM to train and test in patient-level.}
    \scalebox{0.8}{
    \begin{tabular}{cc|ccc|ccc}
        \multicolumn{3}{c}{~} & FFPE & \multicolumn{2}{c}{~} & Frozen & \\
        \hline
        & & Accuracy & Precision & Recall & Accuracy & Precision & Recall \\
        \hline
        &Single training & 0.71 & 0.72 & 0.71 & 0.70 & 0.71 & 0.70 \\
        \cline{2-8}
        & Mixed training & 0.70 & 0.72 & 0.71 & 0.68 & 0.67 & 0.68 \\
        \cline{2-8}
        & MCL & \textbf{0.76} & \textbf{0.77} & \textbf{0.76} & \textbf{0.74} & \textbf{0.73} & \textbf{0.74} \\
        \cline{2-8}
        &ImageNet + A-MIL & 0.72 & 0.72 & 0.72 & 0.70 & 0.68 & 0.70 \\
        \cline{2-8}
        &ImageNet + TransMIL & 0.71 & 0.72 & 0.71 & 0.74 & 0.73 & 0.74 \\
        \cline{2-8}
        &ImageNet + CLAM & 0.71 & 0.72 & 0.71 & 0.69 & 0.70 & 0.69 \\
        \cline{2-8}
        & single training+A-MIL & 0.74 & 0.75 & 0.74 & 0.74 & 0.73 & 0.74 \\
        \cline{2-8}
         & single training+TransMIL & 0.69 & 0.70 & 0.73 & 0.73 & 0.72 & 0.73 \\
        \cline{2-8}
        & single training+CLAM & 0.75 & 0.76 & 0.75 & 0.73 & 0.72 & 0.73 \\
        \cline{2-8}
        & Mixed training+A-MIL & 0.73 & 0.73 & 0.73 & 0.70 & 0.69 & 0.70 \\
        \cline{2-8}
        & Mixed training+TransMIL & 0.70 & 0.70 & 0.70 & 0.72 & 0.73 & 0.72 \\
        \cline{2-8}
        & Mixed training+CLAM & 0.75 & 0.75 & 0.75 & 0.71 & 0.69 & 0.71 \\
        \cline{2-8}
        & MCL+A-MIL & 0.78 & 0.79 & 0.78 & 0.75 & 0.74 & 0.75 \\
        \cline{2-8}
        & MCL+TransMIL & 0.77 & 0.77 & 0.77 & 0.74 & 0.75 & 0.74 \\
        \cline{2-8}
        & MCL+CLAM & \textbf{0.79} & \textbf{0.80} & \textbf{0.79} & \textbf{0.75} & \textbf{0.75} & \textbf{0.75} \\
        \cline{2-8}
        \hline
    \end{tabular}}
\end{table}

\section{Results}
\subsection{Quantitative Results}
For comparison, the evaluation metrics on the CNN backbone with single input training, mixed training, our proposed scheme, SOTAs and combinations of different learning schemes and SOTAs are shown in Table 1. We chose EfficientNet-B0 to evaluate the model performance in these experiments. We observe that our proposed learning scheme outperforms the single and mixed training on the given backbone. These results suggest that the performance of the single training could be limited by the information from each single specific modality, while the mixed training may not efficiently obtain complementary information from the batch-size learning. After being applied to the different learning schemes on SOTAs, our proposed scheme demonstrates the capability of disentangling mutual information from multi-modality and reducing intra-class variance in the latent vector space, which could increase the SOTAs performance at the instance and patient-level.

To further demonstrate that our proposed NMC loss and low-rank loss could fulfill a better representation disentanglement at intra- and inter-patient levels, we compared it with other contrastive loss functions as shown in Table. 2. We observe that the NMC loss provides superior performance, benefiting the representation disentanglement. In comparison, KL-loss fails to consider the distance between positive and negative samples, which might lead to worse performance than other loss functions. As for the marginal triplet and NT-logistic loss, they are measured using the absolute similarity of the positive and negative samples. Although using the relative similarity may help the network optimize the balance between separating the samples of different classes in NT-Xent loss, $l2$ normalization on a single modality can not match differences between modalities. Therefore, the results of NMC-loss illustrate the advantage of layer normalization. Moreover, the low-rank loss shows capacity of the discriminative information disentanglement at intra- and inter-patient levels, compared with other contrastive loss. The combination of the NMC-loss and LR loss achieves best performance. From Table.3, the models in different temperature hyper-parameter have stable accuracy and show the consistent robustness in this task. In our experiments, the training will collapse if temperature is smaller than 0.05.

\begin{table}[htbp]
    \centering
    \caption{Comparison with different contrastive loss functions}
    \scalebox{0.8}{
    \begin{tabular}{ccccc|ccc}
        \multicolumn{2}{c}{~} & FFPE & \multicolumn{2}{c}{~} & Frozen & \\
        \hline
        & & Accuracy & Precision & Recall & Accuracy & Precision & Recall \\
        \hline
         & KL-loss & 0.71 & 0.70 & 0.68 & 0.71 & 0.68 & 0.69 \\
        \cline{2-8}
        & Marginal triplet loss & 0.74 & 0.74 & 0.74 & 0.72 & 0.72 & 0.72 \\
        \cline{2-8}
        & NT-Logistic loss & 0.74 & 0.74 & 0.72 & 0.71 & 0.68 & 0.70 \\
        \cline{2-8}
        & NT-Xent & 0.71 & 0.71 & 0.71 & 0.73 & 0.73 & 0.73 \\
        \cline{2-8}
        & AMC-loss & 0.75 & 0.76 & 0.75 & 0.70 & 0.70 & 0.70 \\
        \cline{2-8}
        & LR loss & 0.75 & 0.77 & 0.75 & 0.72 & 0.73 & 0.72 \\
        \cline{2-8}
        & NMC-loss & 0.75 & 0.75 & 0.75 & 0.72 & 0.71 & 0.72 \\
        \cline{2-8}
        & NMC-loss + LR Loss& \textbf{0.76} & \textbf{0.77} & \textbf{0.76} & \textbf{0.74} & \textbf{0.73} & \textbf{0.74} \\
        \cline{2-8}
        \hline
    \end{tabular}}
\end{table}

\subsection{Visualization}
\begin{figure*}[ht]
    \centering
    \includegraphics[scale=0.4]{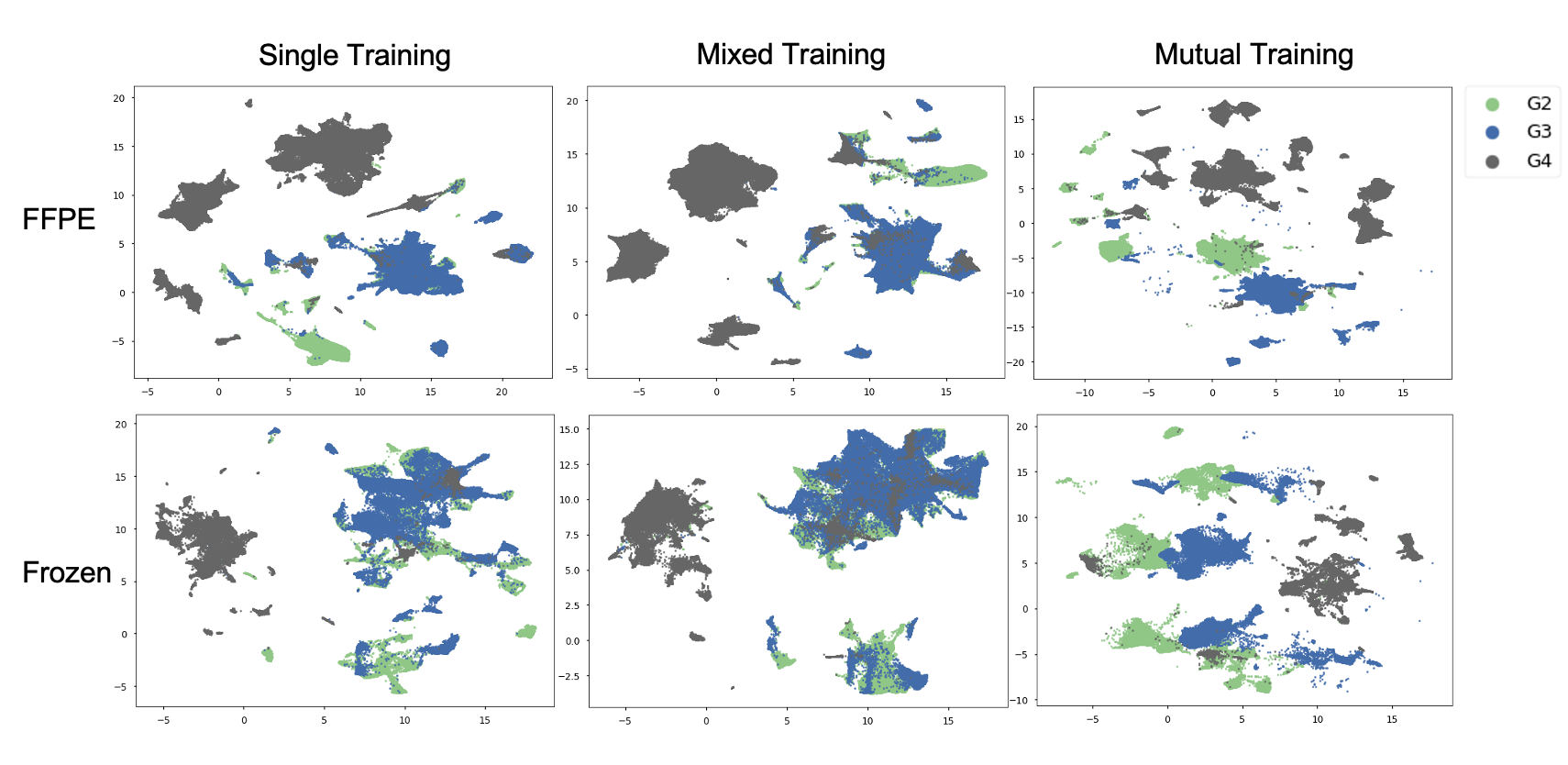}
    \caption{Mutual learning scheme interpretability in brain tumor grading. Each point is from reducing the dimension of the latent vector by UMAP method.} 
\end{figure*}

\begin{figure*}[ht]
    \centering
    \includegraphics[scale=0.3]{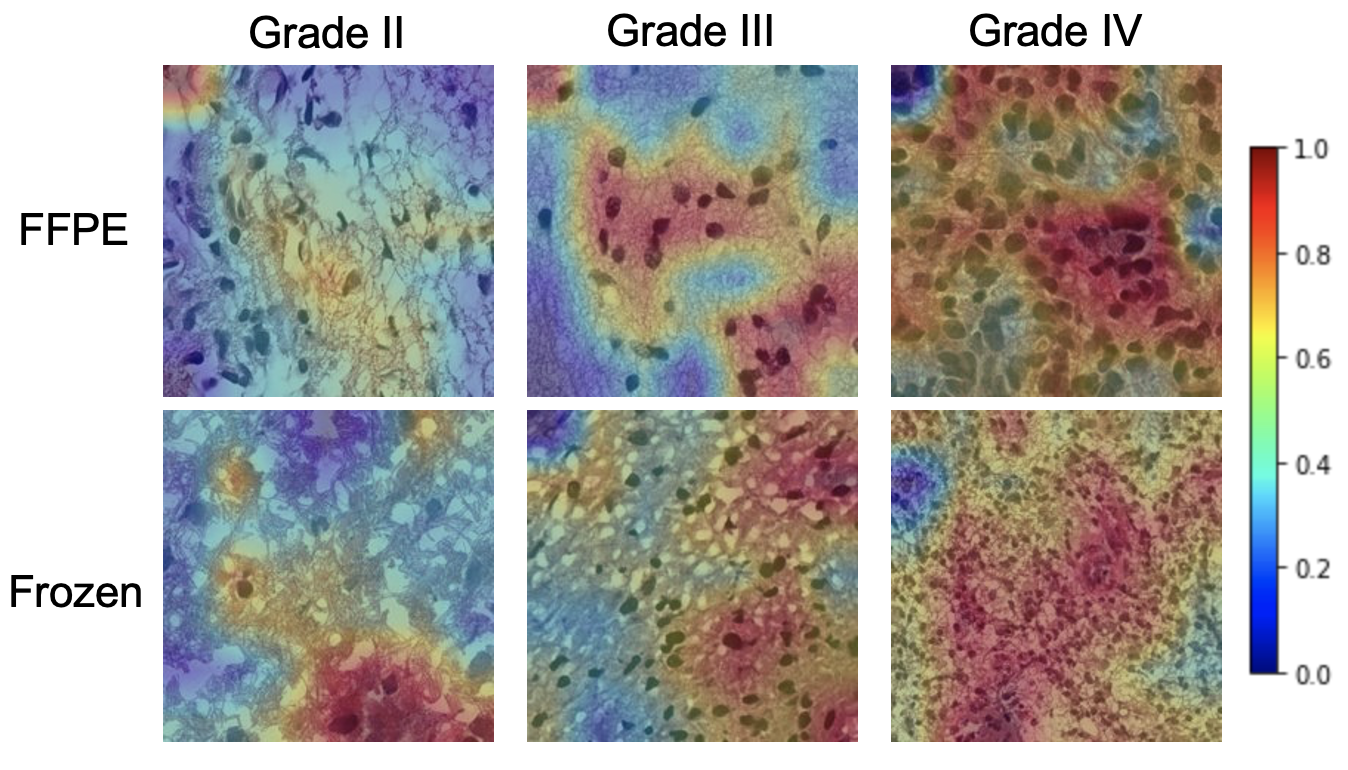}
    \caption{Image-level predicted heatmap of three tumor grades by mutual learning. Left to right: Grade II, III, and IV). The FFPE and Frozen images of each grade are from the same patients. The color bar indicates the estimated level of attention on the region.} 
\end{figure*}

To understand how our proposed scheme leverages representation disentangling in predicting tumor grades, we visualized the latent vector from single training, mixed training and our proposed scheme on FFPE and frozen section images, using the Uniform Manifold Approximation and Projection (UMAP) method \cite{umap}. As shown in Fig. 3, the latent vectors are obtained from the CNN extractors. The results show that our proposed scheme could promote multi-modality and discriminative representation disentanglement, which may help the latent vector on the classification head preserve more helpful information from the same tumor grade, demonstrating closer distribution in the feature space.

\begin{wraptable}[6]{r}{0.4\textwidth}
    \caption{Sensitivity (accuracy) of temperature hyper-parameter}
    \scalebox{0.8}{
    \begin{tabular}{cccccc}
        \hline
        Temperature & 1 & 0.5 & 0.1 & 0.05 \\
        \hline
        FFPE & 0.75 & 0.76 & 0.75 & 0.72 \\
        Frozen & 0.70 & 0.74 & 0.74 & 0.72 \\
        \hline
    \end{tabular}}
\end{wraptable}

The qualitative performance of the randomly selected statistical features from our proposed scheme is illustrated in Fig. 4. These salient maps are generated by classification activation maps (CAMs) \cite{cams}. The examples in Fig. 3 show that the trained model by our proposed scheme can focus on the proliferation region, which helps efficiently detect tissue morphology from the WSI.

\section{Conclusions}
In this paper, we propose a mutual low-rank contrastive clustering learning scheme to improve the performance of representation disentangling on whole slide images for tumor grading. We first develop a mutual learning scheme to extract relevant image representations by integrating FFPE and frozen sections with complementary information. Next, We design an NMC loss, which could promote multi-modality representation disentangling within the same sphere. To further achieve discriminative representation disentanglement on the intra- and inter-patient levels, we conduct a low-rank loss. We demonstrate that our learning scheme could mitigate the drawbacks of the training schemes based on a single modality or mixed modalities and extracted features can further boost the performance of classical attention-based MIL methods. Our other experiments show that our loss outperforms other typical metrics loss. The proposed learning scheme may be generalized to tumor grading of different MIL systems and other cross-modality learning tasks.



%
%
%

\bibliography{egbib}
\end{document}